\documentclass[12pt,osajnl2,preprint,showpacs]{revtex4}
\usepackage[draft]{hyperref}
\usepackage{amsmath,amssymb,graphicx,epsfig,color,bbm}

\def\nn{\nonumber}

\begin{document}

\title{Quantum correlation of an optically controlled quantum system}

\author{Ching-Kit Chan}
\affiliation{Department of Physics, Center for Advanced Nanoscience, University of California San Diego, La Jolla, California 92093-0319, USA}
\author{L. J. Sham}
\email{lsham@ucsd.edu}
\affiliation{Department of Physics, Center for Advanced Nanoscience, University of California San Diego, La Jolla, California 92093-0319, USA}

\begin{abstract} A precise time-dependent control of a quantum system relies on an accurate account of the quantum interference among the system, the control and the environment. A diagrammatic technique has been recently developed to precisely calculate this quantum correlation for a fast multimode coherent photon control against slow relaxation, valid for both Markovian and non-Markovian systems. We review this formalism in comparison with the existing approximate theories and extend it to cases with controls by photon state other than the coherent state.
\end{abstract}


\maketitle

\section{Introduction}\label{sec_introduction}

The study of an open quantum system aims at understanding the underlying physical processes between a quantum system and the environment and at giving a correct account of the system noise due to the environment. Conventional theories usually assume a weak system-environment interaction. This assumption is no longer valid for a controlled quantum system, where the control, as a part of the environment, strongly interacts with the system. Quantum interference between the system and the control and decoherence environment can modify the noise in the quantum system. Given the recent advances in the experimental ultrafast optical techniques, an accurate theory that captures such an quantum interference effect is needed in the regime $t_0 \ll T_2$, where $t_0$ and $T_2$ are the operation and decoherence times, respectively. Also, the stringent error threshold requirement ($\epsilon \sim 10^{-4}$) demanded by fault tolerant quantum computation \cite{aliferis09} is another motivation for a high precision theory for open quantum systems.

The master equation (ME) approach has been successful for Markovian open quantum systems \cite{gardiner00}, where the dynamics of the system is history independent. For such a Markovian system under a coherent control, the corresponding MEs are equivalent to the optical Bloch equations \cite{berman11}, where the control and dissipation terms are additive in the equations. However, for systems with a structured environment \cite{lambropoulos00}, e.g. photonic crystal materials \cite{lodahl04} and nanocavity systems \cite{noda00, yoshie04}, the dynamics are non-Markovian and the optical Bloch theory is inapplicable. In fact, because of the memory effect of these systems, interferences exist between the control and the environment and the validity of the additive form of the ME is questionable. Therefore, the challenge remains open in non-Markovian systems.

A precise diagrammatic technique has been formulated recently to solve the dynamics of a two level system (TLS) under a time-dependent coherent photon control through the multimode Jaynes-Cummings (JC) interaction in the regime $t_0 \ll T_2$ \cite{chan11}. It takes into full account of the quantum interference among the TLS, the control and the electromagnetic vacuum and is applicable for both Markovian and non-Markovian situations. The idea is to construct the time evolution of the TLS in terms of a time loop propagation, in the same spirit as the Keldysh non-equilibrium Green's function technique \cite{keldysh65}. Under this construction, the underlying physical processes become transparent: dissipations originate from the contraction between photons at different times; coherent Rabi motions are due to propagators dressed by the control photons; interferences between these two processes result in the control noise. For $t_0 \ll T_2$, only diagrams with a small number of contractions are important and they can be summed exactly, allowing a precise calculation of the controlled dynamics of the TLS in this time domain within a given error bound. Based on a simple non-Markovian model, it has been shown that this diagrammatic formalism captured quantum correlations of the entire system that was missed in conventional ME approximations even in the small time limit \cite{chan11}. The error of the ME approximations is comparable to that of the classical approximation.

This diagrammatic approach is not restricted to the case where the photon control is a Glauber state. The control can formally be extended to an arbitrary photon state. To illustrate this, we extend the theory to the cases of a multimode squeezed coherent state and a multimode photon number state. Comparisons with the single mode results indicate the accuracy of these diagrammatic solutions.

The paper is organized as follows. In Section~\ref{sec_diagrammatic approach}, we review the diagrammatic formalism for a TLS under a coherent control and provide rules for the diagrammatic construction. Section~\ref{sec_comparison} performs an explicit comparison with the standard ME approaches. We also demonstrate the ability of the diagrammatic approach to systematically improve the accuracy by including higher order diagrams. Section~ \ref{sec_non-coherent control} extends the discussion to non-Glauber photon controls. We conclude in Section~\ref{sec_conclusion}.

\section{TLS under coherent control}\label{sec_diagrammatic approach}

The central idea of the diagrammatic technique is first to express the time evolution of the composite system in terms of the spin and photon operators, then to evaluate the transition matrix in terms of photon operators, which are finally treated diagrammatically using the Wick's expansion. The problem of a time-dependent controlled TLS is governed by the multimode JC Hamiltonian under a multimode coherent state $\left| \boldsymbol{\alpha} \right\rangle= \left| \alpha_{k_1},\alpha_{k_2},\ldots \right\rangle$. Consider first for simplicity, an initial wavefunction of the product form $|\Psi(0)\rangle =\left[\sum_s c_{s} |s\rangle \right]|\boldsymbol{\alpha}\rangle $, where $s=\pm$ gives the spin state. The system evolves as $|\Psi(t)\rangle = U(t) |\Psi(0)\rangle  $, where the evolution operator in the interaction picture is given by:
\begin{eqnarray}
U(t) &=& T \exp \Big[ -i \int_0^t dt' V(t') \Big],
\label{eq-v}
\end{eqnarray}
and the multimode JC interaction is:
\begin{eqnarray}
V(t_l) &=& \sigma_+ A_l + \sigma_- A_l^\dagger, \\
\text{with~~}
A_{l}&=&\sum_{k}g_{k}a_{k}e^{i(\omega_0-\omega_k) t_{l}}.
\label{eq-u4}
\end{eqnarray}
with $\omega_0$ the energy splitting of the TLS, $\omega_k$ the photon frequency and $g_k$ the TLS-photon coupling strength. The interaction on the initial coherent photon state can be viewed as the action of a control pulse, consisting of a modulation of the resonant frequency mode with the coherent state amplitude satisfying:
\begin{eqnarray}
\sum_{k}\left(g_{k}e^{-i\omega_{k}t}\alpha_{k}+c.c\right) = \frac{\Omega(t)}{2}e^{-i(\omega_{0}t+\phi)}+\text{c.c.},
\label{eq_E(t)}
\end{eqnarray}
where $\alpha_k$ is the eigenvalue of $a_k$ on $\left| \boldsymbol{\alpha} \right\rangle$, $\Omega(t)$ and $\phi$ are the envelope function and phase of the pulse, respectively. Note that we have made the rotating wave approximation which is valid when $\omega_0 t\gg g|\alpha|t \sim O(1)$ \cite{chan11}. The physical quantity of interest is the transformation matrix:
\begin{eqnarray}
p_{s_f,s_f';s,s'}(t,\boldsymbol{\alpha}) &=& \langle\boldsymbol{\alpha} | \langle s'| U^{\dagger}(t) |s_f' \rangle \langle s_f| U(t) |s\rangle |\boldsymbol{\alpha} \rangle \nn \\
&& \times \ e^{i(s_f'1-s_f1)\omega_0 t/2},
\label{eq_transformation_matrix}
\end{eqnarray}
from which we can obtain the reduced density matrix of the TLS at that:
\begin{eqnarray}
P_{s_f,s_f'}(t) =  \sum_{s,s'} c_{s}c_{s'}^* p_{s_f,s_f';s,s'}(t,\boldsymbol{\alpha}).
\label{eq_reduced density matrix}
\end{eqnarray}
It is straightforward to generalize the above formulae to an initial state which is a linear combination of the product states.

In contrast to the ME methodology which traces out the photonic environment in the beginning, we evaluate the TLS transition by the spin operators in the transformation matrix first, leaving a sequence of the corresponding photon components. An example is given by:
\begin{eqnarray}
&&p_{++;++}(t)\nn \\ &=& \sum_{n,n'=0}^\infty (-1)^{n+n'} \int_0^t D^{2n} t \int_0^t D^{2n'} t \nn \\
&\times &\left\langle \boldsymbol{\alpha} \right|\left(A_{1'} A_{2'}^\dagger ... A_{2n'-1} A_{2n'}^\dagger\right) \left( A_{2n} A_{2n-1}^\dagger ... A_{2} A_{1}^\dagger\right) \left| \boldsymbol{\alpha} \right\rangle, \nn \\
\label{eq_infinite series}
\end{eqnarray}
where $\int_0^t D^{n} t=\int_0^t dt_n ...\int_0^{t_3} dt_2\int_0^{t_2} dt_1$. In the following, we will focus on this expression for illustration. Other transformation matrices can be obtained similarly (see \cite{chan11}). Therefore, the task is now reduced to compute expectation values of the photon operators for the initial multimode coherent state.

\begin{figure}[htb]
\centering
\includegraphics[width=1.0\columnwidth]{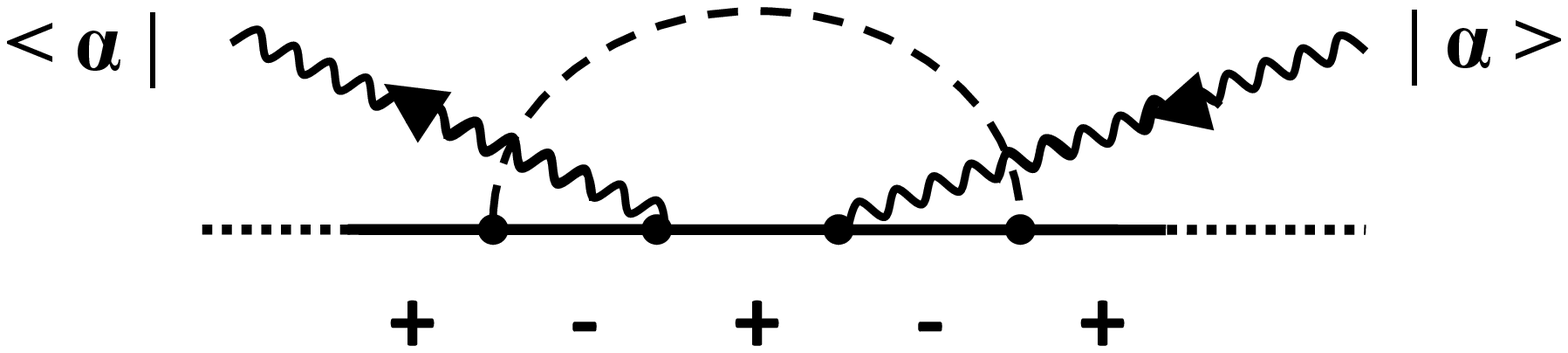}
\caption{An illustration of the TLS-multiphoton process conditioned on a multimode coherent state.  The $\pm$ sign labels the spin state of the TLS. Contractions among photons (dashed line) correspond to decoherence, while the normal-ordered control photons (wavy lines) is responsible for the Rabi motion of the TLS.
 diagram0.eps}
\label{diagram0}
\end{figure}

Eq.~(\ref{eq_infinite series}) is exact up to now and all the quantum correlations of the whole system remain intact. The underlying physical processes can be easily identified using a diagrammatic representation as in Fig.~\ref{diagram0}. We shall see in the next subsection how different roles of the photons affect the interaction dynamics, and how this infinite series expression can be largely simplified and summed to a desirable accuracy using $t/T_2$ as a small parameter. No stochastic assumption has been made. We emphasize that this approach is fundamentally different from the ME method which usually assumes a stationary environment and thus does not account for the interference effect between the control and the environment of a quantum system.

\subsection{Diagrammatic solution}\label{sec_diagrammatic solution}

It is instructive to describe Eq.~(\ref{eq_infinite series}) as time loop integrals of photon operators propagating forward and backward between time $0$ and $t$, and depict Fig.~{\ref{diagram0}} in a bubble form. Diagrammatic series is constructed using the Wick's theorem \cite{wick50}, which expands the product of photon operators in terms of the normal ordered form and possible contractions between two photon operators. By doing so, the expectation value in Eq.~(\ref{eq_infinite series}) can be computed easily by the replacement $a_k \rightarrow \alpha_k$ and $a_k^\dagger\rightarrow \alpha_k^*$ for normal ordered photons. Physically, the normal ordered photons coming from the multimode coherent state are responsible for the coherent Rabi motion of the TLS; while the contraction between photons, being independent of the control, corresponds to vacuum relaxation of the TLS.

\begin{figure}[htb]
\centering
\includegraphics[width=1.0\columnwidth]{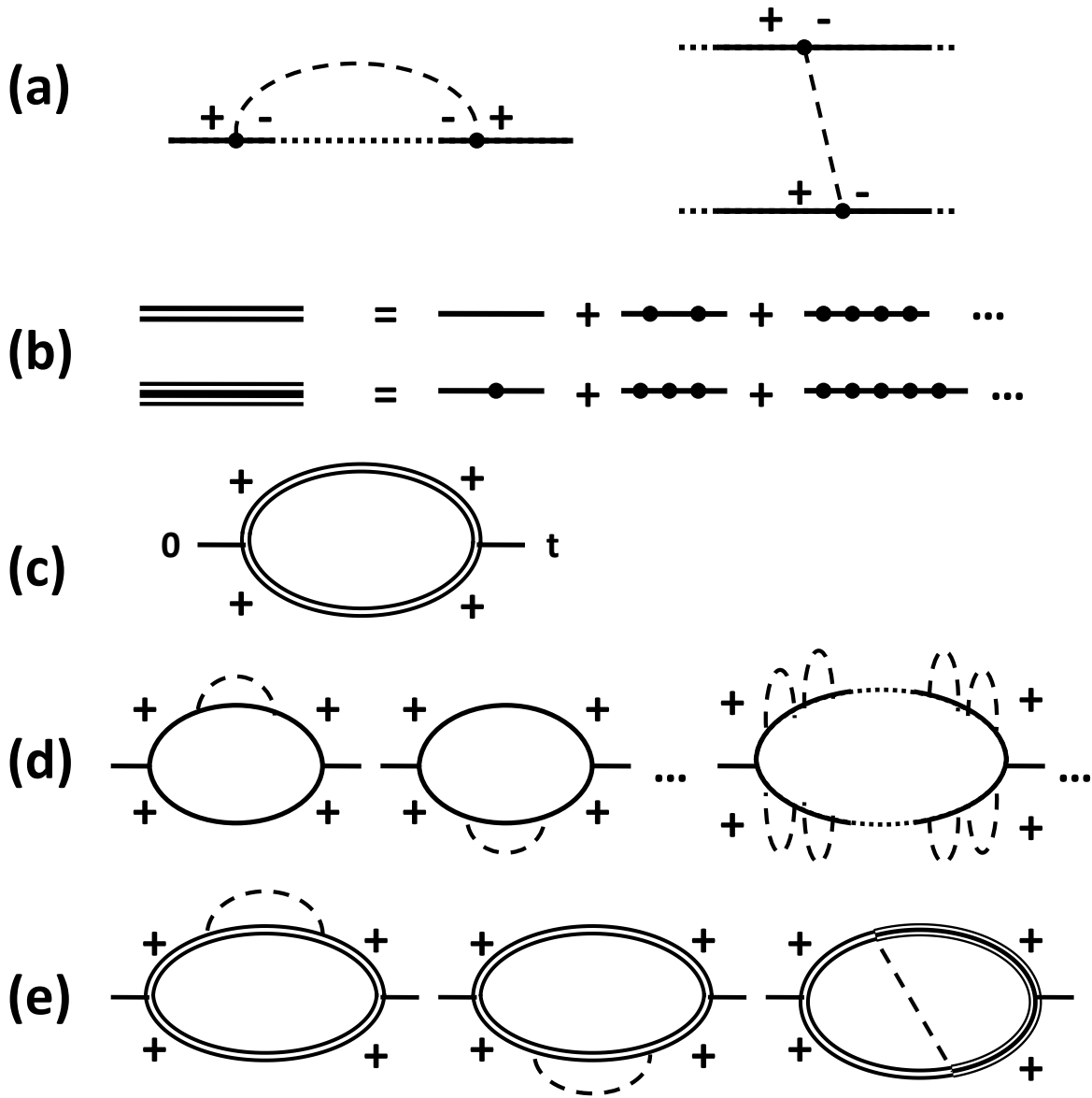}
\caption{Diagrammatic representations of the Wick's expansion of the transformation matrix $p_{s_f,s_f';s,s'}(t,\{\alpha\})$ for a coherent control. (a) The two possible contractions between photons at different times. (b) Two types of dressed line that represent a sum of even or odd numbers of control photons interacting with the TLS. (c) A dressed diagram without any contraction represents the classical Rabi solution of $p_{++;++}$ from time $0$ to $t$. (d) Vacuum relaxation is represented by undressed diagrams with contractions only. (e) The leading contribution of control noise to $p_{++;++}$ comes from three dressed diagrams with only one contraction.
 diagram1.eps}
\label{diagram1}
\end{figure}

The diagrammatic structure consists of two ingredients: the contraction between photons and dressed lines due to the coherent control. The contraction function is given by:
\begin{eqnarray}
\langle A_i A_{j}^\dagger \rangle &=&  K(t_i-t_j)= \sum_k g_k^2 e^{i (\omega_0-\omega_k) (t_i - t_j)},
\label{eq_contraction}
\end{eqnarray}
and is depicted in Fig.~\ref{diagram1}a. Here, $t_i$ and $t_j$ can be on the same or opposite time lines. Note that the contraction only depends on the photon density of state (DOS) $\rho(\omega)=\sum_{k}g_{k}^{2}\delta(\omega-\omega_{k})$. One can show that each contraction line $\sim \int \int dt_i dt_j K(t_i-t_j)\sim O[(t/T_2)^\gamma] $, where $1 \leq \gamma \leq 2$ depends on the DOS \cite{chan11}. Therefore, the Wick's expansion in terms of contraction is automatically a perturbative series of the small parameter $(t/T_2)^\gamma$. On the other hand, the dressed lines lead to Rabi oscillations of the TLS and are given by \cite{chan11}:
\begin{eqnarray}
D_{e} (t , t') &=& \cos\left( \frac{\mathcal{A}(t)-\mathcal{A}(t')}{2}\right) \Theta(t-t'),\nn \\
D_{o} (t , t') &=& (\pm i) (e^{\pm i\phi}) \sin\left( \frac{\mathcal{A}(t)-\mathcal{A}(t')}{2}\right) \Theta(t-t')  ,
\label{eq_dressed_funcyion1}
\end{eqnarray}
corresponding to the double and triple line depicted in Fig.~\ref{diagram1}b. A dressed line segment between the same (opposite) spin state corresponds to a double (triple) line. Here, $D_o(t,t')$ picks the factor $+i$ ($-i$), when the triple line is on the upper (lower) time line; and gains a phase $e^{i\phi}$ ($e^{-i\phi}$), if it goes from $-$ to $+$ in the clockwise (anticlockwise) sense. $\mathcal{A}(t)=\int_0^{t'} dt' \Omega(t')$ gives the area of the coherent pulse at time $t$.

The diagrammatic procedure can be summarized as follows: (i) denote the initial and final state of the transformation matrix on a loop diagram, where the lower and upper arcs represent forward and backward time evolutions; (ii) draw all possible contractions according to Fig.~\ref{diagram1}a; and (iii) dress the bare lines according to the spin flips by the photons. Each diagram can be computed straightforwardly by integrating over all time variables at the vertices of each contraction line. The vertex picks up a factor of $i$ ($-i$) when it is on the upper (lower) time line.

The diagrammatic structure allows an easy identification of the underlying coherent and dissipative processes. For instance, Fig.~\ref{diagram1}c shows the diagram with no contraction for $p_{++;++}(t)$. Following the above description, this diagram simply gives the expected coherent Rabi solution $p_{++;++}^{(0)}(t)=\cos^2\left[\mathcal{A}(t)/2 \right]$. On the other hand, in the absence of control, the vacuum relaxation is represented by a series of undressed diagrams that involves contractions only, as can be seen in Fig.~\ref{diagram1}d. In the regime $t \ll T_2$, the first two diagrams in Fig.~\ref{diagram1}d results in a decoherence of $\sim O[(t/T_2)^\gamma]$.

The leading order contribution to the control noise problem is represented by the three diagrams in Fig.~\ref{diagram1}e. Interference between the control (dressed line) and decoherence (contraction line) is evident. We stress that these diagrams are first order in decoherence, but infinite order in the coherent interaction. This explains why the diagrammatic approach is a suitable candidate to our control noise problem, in contrast to the ME method that requires a weak system-environment coupling.

\subsection{Application to a general DOS}\label{sec_application}

The field theoretic approach described above is applicable for an arbitrary photon DOS and thus valid for both the Markovian and non-Markovian systems. For a broadband DOS, the contraction function becomes a delta function $K(t_i-t_j)\sim \delta(t_i-t_j)$, meaning that the two ends of a contraction line are squeezed to the same point. In this limit, the correlation time $\tau_c$ is zero and the decoherence is linear in $t/T_2$ (i.e. $\gamma = 1$), corresponding to a Markovian exponential decay. The dynamics of the controlled TLS can be alternatively obtained by the optical Bloch equation and the diagrammatic solution shows quantitative agreement with it in the regime $t \ll T_2$ \cite{chan11}.

Extension to the non-Markovian regime can be performed simply by modifying the photon DOS and thus the contraction function. The correlation time $\tau_c$ is characterized by the bandwidth of the DOS. The contraction function is no longer a delta function, leading to non-exponential decoherence $\sim O[(t/T_2)^\gamma]$, where $1 < \gamma \leq 2$. Furthermore, there is a transition from the non-exponential to exponential decay as $t$ becomes comparable to $\tau_c$. All these features remain in the presence of a coherent control \cite{chan11}. In the following section, we will investigate the accuracy of the diagrammatic scheme in comparison with existing ME methods.

\section{Comparison with ME approaches}\label{sec_comparison}

Conventional master equation (ME) assumes an additive form \cite{prataviera99}:
\begin{eqnarray}
\frac{d}{dt}\rho_s(t)=-i[H_{c}(t),\rho_s(t)]+\int_0^t dt' \hat L(t-t') \rho_s(t'),
\label{eq_additive ME}
\end{eqnarray}
where $H_c(t)$ describes a control Hamiltonian that only acts on the system and $\hat L(t-t')$ is a suitably chosen superoperator that only accounts for the dissipative environment. It is assumed that these two terms are independent on each other. In this section, we examine the validity of this kind of ME against the field theoretic solution in Section~\ref{sec_diagrammatic approach}.

\subsection{Second order TLS-environment interaction}\label{sec_higher order}

We consider two commonly adopted ME approaches based on the projective operator techniques, namely the Nakajima-Zwanzig (NZ) and the time-convolutionless (TCL) methods \cite{breuer99, breuer02}. We now show that they belong to the additive ME in Eq.~(\ref{eq_additive ME}). The second order NZ ME is given by:
\begin{eqnarray}
&&\frac{d}{dt}\rho_s^\text{NZ}(t) \nn \\
&= &-i \text{Tr}_R\left[ V(t),\rho_s^\text{NZ}(t)\otimes \rho_R  \right]\nn \\
&&+\int_0^t dt' \text{Tr}_R \left[ V(t), \text{Tr}_R \left[V(t'),\rho_s^\text{NZ} (t')\otimes \rho_R \right]\otimes \rho_R  \right]\nn \\
&&-\int_0^t dt' \text{Tr}_R \left[ V(t),\left[V(t'),\rho_s^{NZ} (t')\otimes \rho_R \right]  \right],
\label{eq_NZ}
\end{eqnarray}
where $\rho_s(t)$ is the reduced density matrix of the TLS, $\rho_R = |\boldsymbol{\alpha}\rangle \langle\boldsymbol{\alpha}|$ denotes the initial multimode photon density matrix and has no time dependence. The non-vanishing first and second terms are due to fact that $\text{Tr}_R [V(t) \otimes \rho_R] \neq 0$. Using the multimode JC interaction in Eq.~(\ref{eq-v}) and the relation in Eq.~(\ref{eq_E(t)}), we trace out the photon reservoir and arrive at the NZ ME:
\begin{eqnarray}
&&\frac{d}{dt}\rho_s^\text{NZ}(t) \nn \\
&= &-i \frac{\Omega(t)}{2}\left[ \sigma_+ e^{i\phi}+\sigma_- e^{-i\phi}, \rho_s^\text{NZ}(t) \right]\nn \\
&&-\int_0^t dt' K(t-t') \left[ \sigma_+\sigma_- \rho_s^\text{NZ}(t')-\sigma_-\rho_s^\text{NZ}(t')\sigma_+ +h.c.\right], \nn \\
\label{eq_NZ2}
\end{eqnarray}
where $K(t-t')$ is just the contraction function in Eq.~(\ref{eq_contraction}). The second order TCL ME can be obtained simply by the replacement $\rho_s^{TCL} (t') \rightarrow \rho_s^{TCL} (t)$ in the integrand.

\begin{figure}[htb]
\centering
\includegraphics[width=1.0\columnwidth]{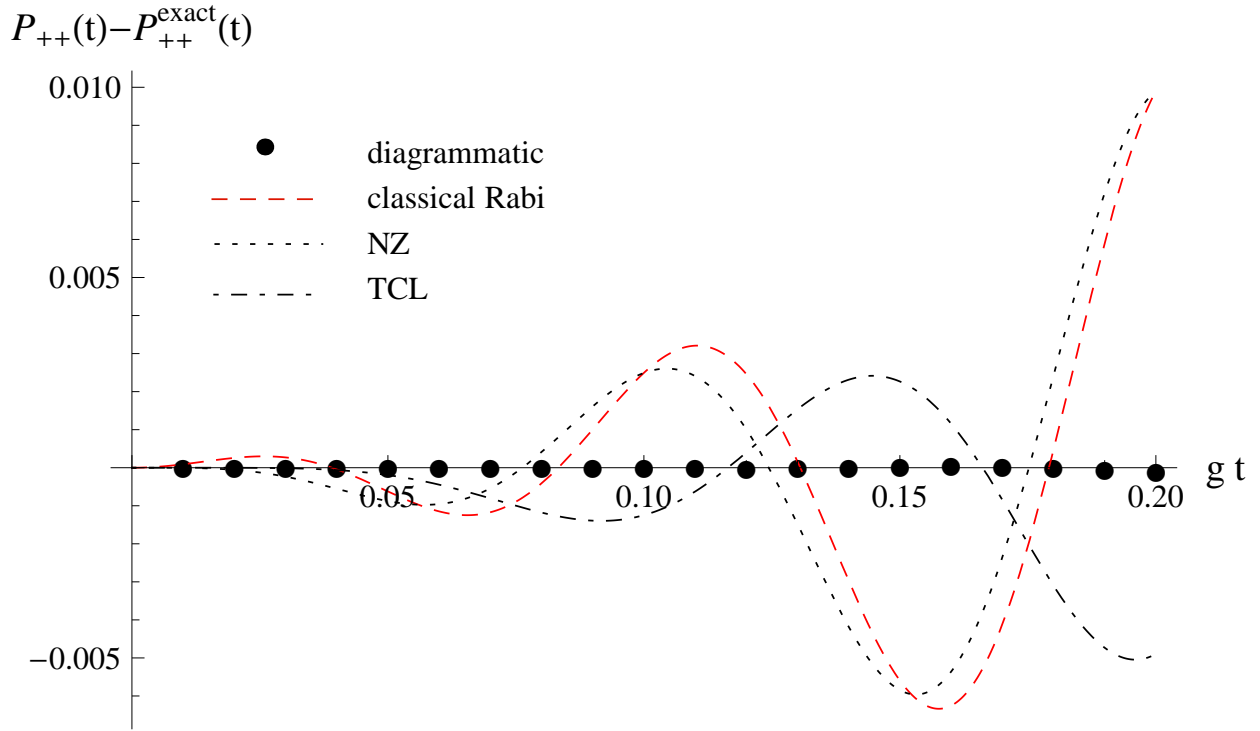}
\caption{(color online) A comparison between the diagrammatic methods and the ME approaches presented in \cite{chan11}, using a coherently driven single mode JC system with $\bar n = 100 \pi^2$, corresponding to a $4\pi$ rotation at $gt=0.2$. The magnitudes of errors of the ME approaches and the classical Rabi solution are comparable even in the small time regime.
comparison1.eps}
\label{comparison1}
\end{figure}

It is clear from Eq.~(\ref{eq_NZ2}) that the NZ ME and TCL ME belong to the class of additive ME in Eq.~(\ref{eq_additive ME}), where the corresponding superoperators $\hat L (t-t')$ do not depend on the control, and thus both of them have neglected the quantum interference effect between the control and dissipation. This can be explicitly illustrated by using a simple non-Markovian system, the single mode JC model, where an exact solution is available. This situation corresponds to a constant contraction function $K(t-t')=g^2$, infinite correlation time $\tau_c=\infty$, $\gamma =2$ and $T_2 = \sqrt{2} /g$. We consider an initially excited TLS under a single mode coherent drive with $\bar n = 100\pi^2$, so that the system undergoes a $4\pi$ rotation when $gt=0.2$. Fig.~\ref{comparison1} compares the difference between different approaches and the exact solution for $P_{++}(t) = p_{++;++}(t)$. Here, the diagrammatic solution, based on the three diagrams in Fig.~\ref{diagram1}e, shows an excellent agreement with the exact solution in the small $gt$ regime. The error of these three diagrams is $O[(gt)^4]$. On the contrary, these two ME methods, expected to be the same and accurate to the second order of $gt$ \cite{breuer99, breuer02}, show errors of the same order of magnitude. The error of the NZ solution is very close to that of the classical Rabi solution ($P_{++}(t) = \cos^2 (g\sqrt {\bar n}t)$) which entirely neglects any decoherence effect; while the error of TCL result shows a phase difference from the NZ and classical solution because of the approximation $\rho_s^{TCL} (t') \rightarrow \rho_s^{TCL} (t)$ in the integrand of the ME.

The problem of these ME approximations stem from the assumptions of a stationary environment, which is inappropriate for a quantum system under a general control. Quantum feedback from the photon environment to the TLS exists and originates from the quantum interference between the control and the environment. In fact, the difference between the field theoretic solution and the additive ME method can be viewed as the difference between the quantum and semiclassical correlations. This difference is not limited to the single mode test above and occurs for a general non-Markovian multimode system.

\subsection{Higher order terms}\label{sec_higher order}

\begin{figure}[htb]
\centering
\includegraphics[width=1.0\columnwidth]{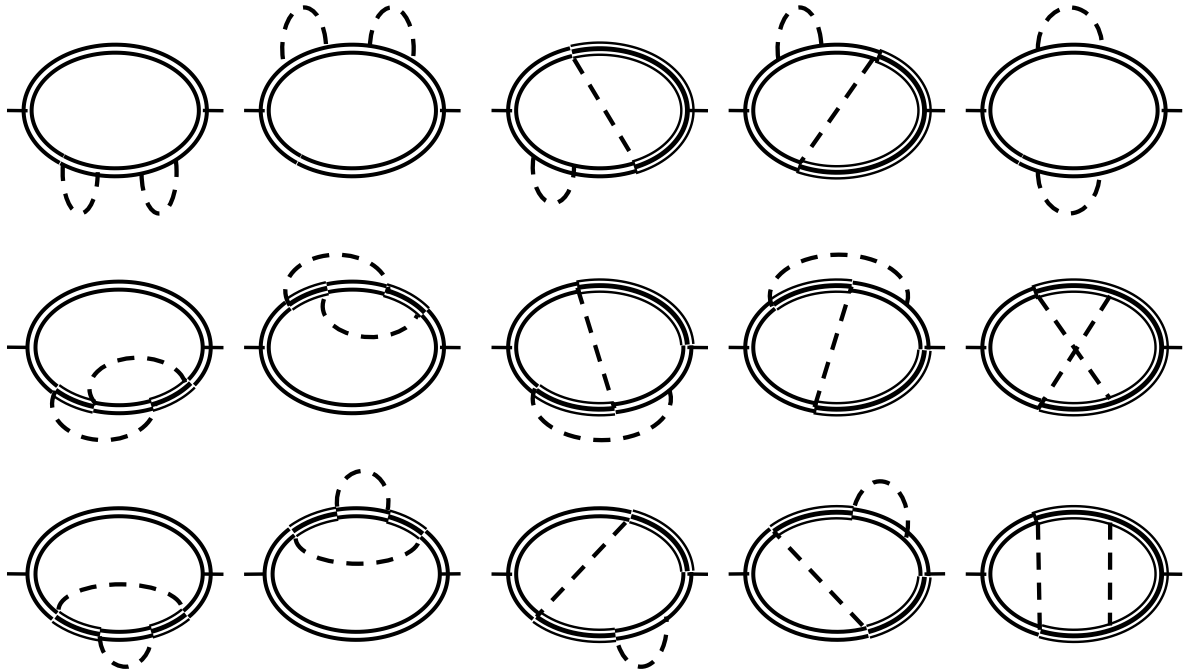}
\caption{Fifteen dressed diagrams with two contractions for $p_{++;++}$ under a coherent control.
diagram2.eps}
\label{diagram2}
\end{figure}

\begin{figure}[htb]
\centering
\includegraphics[width=1.0\columnwidth]{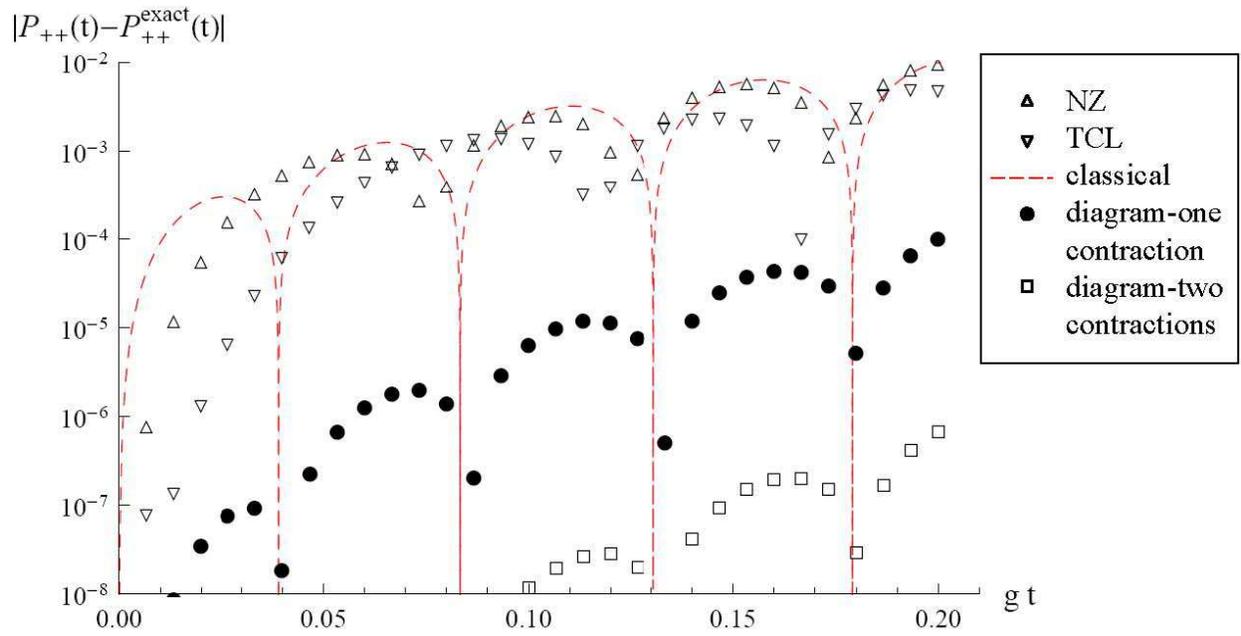}
\caption{(color online) A plot of absolute errors of different methods under the same physical situation as in Fig.~\ref{comparison1}. Diagrammatic solution with $n$ contractions has an error of $O\left[(gt)^{2(n+1)} \right]$.
comparison2legend.eps}
\label{comparison2legend}
\end{figure}

The precision of the diagrammatic solution can be further improved by including higher order diagrams. In general, the inclusion of diagrams with $n$ contractions would result in an error of $O[(t/T_2)^{(n+1)\gamma}]$. Fig.~\ref{diagram2} provides the next order diagrams that possess two contractions for $p_{++;++}(t)$. They can be evaluated by using the same procedure described in Section~\ref{sec_diagrammatic solution}. As an example, Fig.~\ref{comparison2legend} plots the absolute difference of $P_{++}(t)$ between different methods and the exact solution in log scale under the same physical situation as in Fig.~\ref{comparison1}. It is apparent that the diagrammatic solution has a well-controlled error bound in the small time domain.

\section{Control by photon states other than coherent state}\label{sec_non-coherent control}

The field theoretic technique can be formulated similarly when the TLS is initially in the presence of a quantum photon state other than the coherent state. The Wick's expansion is still valid, though the diagrammatic rules have to be modified. Here, we extend this formalism to the multimode squeezed coherent state and the multimode photon number state. Rather than exploring all possible scenarios, the purpose of this section is to demonstrate how diagrammatic solutions can be constructed with different initial photon states.

\subsection{Squeezed coherent state}\label{sec_squeezed state}

A multimode squeezed coherent state takes the form $\left| \boldsymbol{\alpha}, \xi \right\rangle = D( \boldsymbol{\alpha}) S(\xi) \left|0 \right\rangle$, where $D( \boldsymbol{\alpha})$ is the multimode displacement operator and we take the multimode squeezing operator to be $S=\prod_k \exp\left(\xi^* a_{\bar k+k} a_{\bar k-k}-\xi a_{\bar k+k}^\dagger a_{\bar k-k}^\dagger\right)$ \cite{scully97}. $\bar k$ governs on the squeezing mechanism. The problem is to evaluate Eq.~(\ref{eq_infinite series}) using $\left| \boldsymbol{\alpha}, \xi \right\rangle $ instead of $\left| \boldsymbol{\alpha}\right\rangle $.

\begin{figure}[htb]
\centering
\includegraphics[width=1.0\columnwidth]{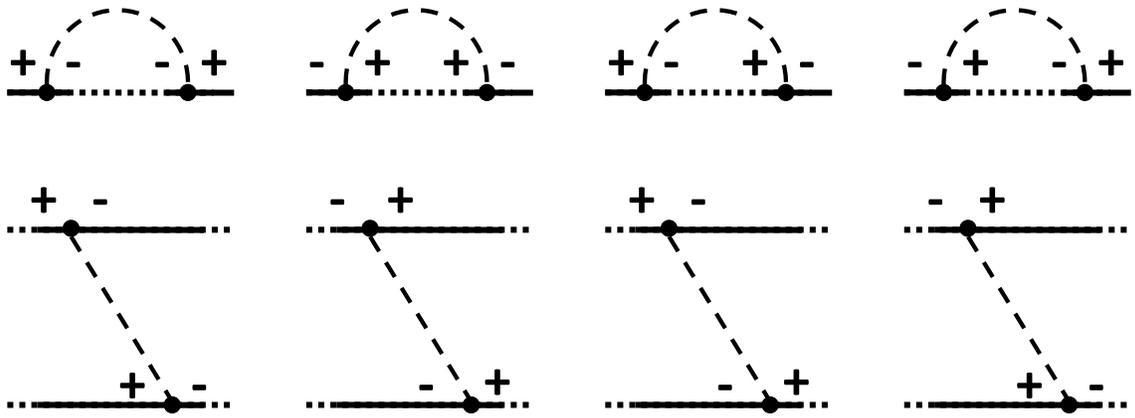}
\caption{Eight possible contractions between photons for a squeezed vacuum.
 diagram3.eps}
\label{diagram3}
\end{figure}

Making use of the squeezing transformation, we can rewrite Eq.~(\ref{eq_infinite series}) by the replacement $A_i \rightarrow \tilde {A_i}=S^\dagger A_i S$, $A_{j}^\dagger \rightarrow \tilde {A}_{j}^\dagger=S^\dagger A_{j}^\dagger S$ and $D( \boldsymbol{\alpha}) \rightarrow \tilde {D}( \boldsymbol{\alpha})=S^\dagger D( \boldsymbol{\alpha}) S$, where each photon operator transforms according to
\begin{eqnarray}
S^\dagger a_k S &=& a_k \cosh r - a_{2\bar k-k}^\dagger e^{i\theta} \sinh r, \nn \\
 S^\dagger a_k^\dagger S &=& a_k^\dagger \cosh r - a_{2\bar k-k} e^{-i\theta} \sinh r,
\label{eq_squeezing transformation}
\end{eqnarray}
and the squeezing parameters satisfy $\xi = r e^{i\theta}$. Now, we can proceed the Wick's expansion as in Section~\ref{sec_diagrammatic solution}. The dressed lines are unaffected as the normal ordered photon operators are just $c$-numbers. As such, the zeroth order diagram remains the same. However, four types of contractions are now possible due to the squeezing:
\begin{eqnarray}
\langle \tilde A_i \tilde A_j^\dagger \rangle &=&  \cosh^2 r \sum_k g_k^2 e^{i(\omega_0 -\omega_k) (t_i - t_j)},\nn \\
\langle \tilde A_i^\dagger \tilde A_j \rangle &=&  \sinh^2 r \sum_k g_k^2 e^{-i(\omega_0 -\omega_k) (t_i - t_j)},\nn \\
\langle \tilde A_i \tilde A_j \rangle &=& -\frac{e^{i\theta}}{2} \sinh(2r) \nn \\
 &&\times \sum_k g_k g_{2\bar k-k} e^{i(\omega_0 -\omega_k) t_i }e^{i(\omega_0 -\omega_{2\bar k-k}) t_j },\nn \\
 \langle \tilde A_i^\dagger \tilde A_j^\dagger \rangle &=& \langle \tilde A_i \tilde A_j \rangle^*,
\label{eq_squeezed contraction}
\end{eqnarray}
and they are depicted in Fig.~\ref{diagram3}. The above dependence of the contraction functions on the squeezing parameter $\theta$ shows a phase sensitive decoherence of the TLS.

\begin{figure}[htb]
\centering
\includegraphics[width=1.0\columnwidth]{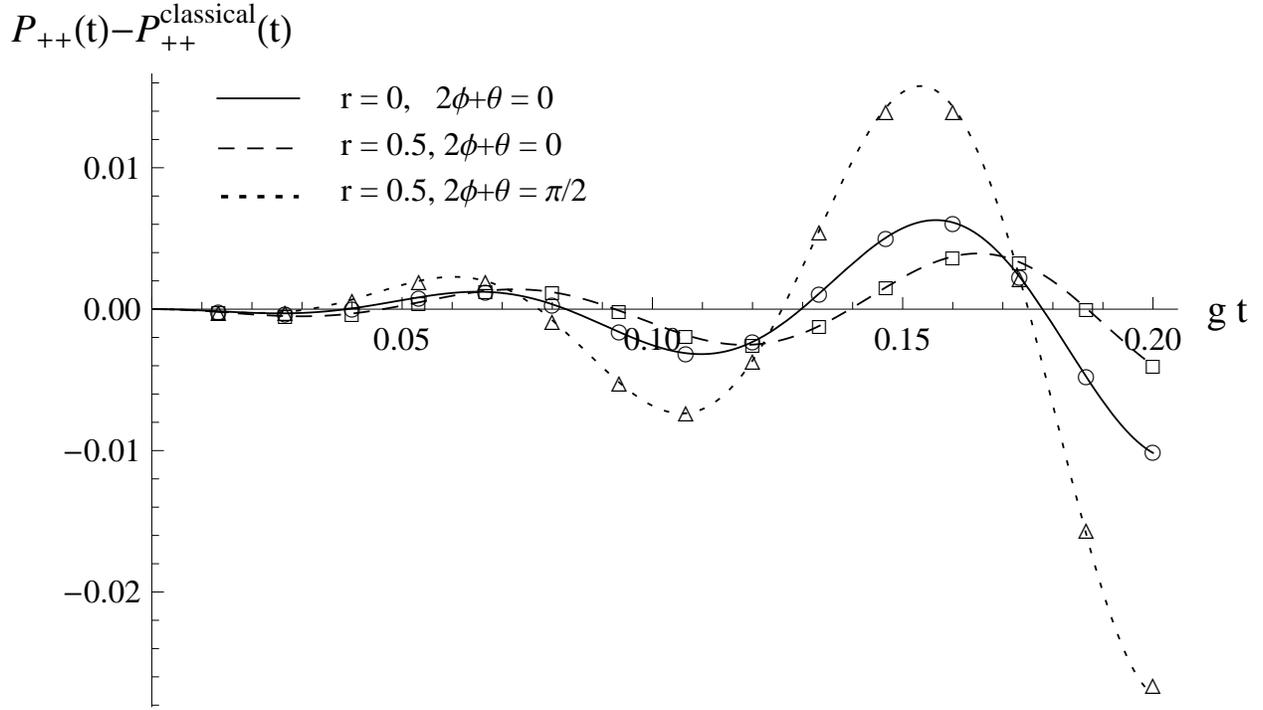}
\caption{$P_{++}(t)-P_{++}^{\text{classical}}(t)$ of an initially exited TLS under a single mode squeezed coherent state with $\bar n = 100 \pi^2$ for various squeezing parameters. $\phi$ is the phase of $\alpha^*$ and $\theta$ is the phase of squeezing. The curves and symbols correspond to the diagrammatic and exact solutions, respectively.
 squeezedstate.eps}
\label{squeezedstate}
\end{figure}

In brief, the diagrammatic rules in Section~\ref{sec_diagrammatic solution} still apply with the exception that the construction of contractions in Fig.~\ref{diagram1}a is replaced by that in Fig.~\ref{diagram3}. The leading order contribution to control noise now consists of twelve diagrams which are similar to the three shown in Fig.~\ref{diagram1}e. To demonstrate the accuracy of the diagrammatic solution, we compare it with an exact solution of a single mode JC system under a squeezed coherent state \cite{milburn84}. Fig.~\ref{squeezedstate} again shows an excellent precision of the diagrammatic solution. Depending on the squeezing parameters, the phase sensitive decoherence of the TLS can be greater or smaller than that without squeezing. We note that since the decoherence time is modified to $\sim 1/(g \cosh r)$, the small time requirement becomes $g \cosh (r) t \ll 1$ for the validity of the perturbative solution.

The above squeezed state analysis works similarly for the more general linearly transformed coherent state $D( \boldsymbol{\alpha}) T \left|0 \right\rangle$, where $T$ is any unitary operator that transforms the photon operators linearly, $T^\dagger a_k T = \sum_{k'} \left (u_{kk'} a_{k'} + v_{kk'} a_{k'}^\dagger\right)$. The diagrammatic solution shows that the coherent motion of the TLS remains the same, while the decoherence can be computed straightforwardly by using the modified contraction functions similar to Eq.~(\ref{eq_squeezed contraction}).

\subsection{Number state}\label{sec_number state}

The dynamics between a TLS and a multimode number state is more complicated. A single photon process can project the photon state to be orthogonal to the initial state. It is a daunting task to keep track of the time evolution of a number state with an arbitrary photon number distribution. With the help of the diagrammatic technique, we shall see below how it is possible to tackle a number state of the form $\left|N, \{0\} \right\rangle$, where $\{0\}$ is a multimode vacuum.

The idea is again to make use of the Wick's expansion of the photon operators in Eq.~(\ref{eq_infinite series}), where the coherent state is replaced by $\left|N, \{0\} \right\rangle$ here. We assume the $N$-mode of the number state has zero detuning. The zeroth order term has all the photon operators being normal ordered, so that each photon operator can only annihilate or create photons in the $N$-mode. Thus, the time integrals in Eq.~(\ref{eq_infinite series}) can be performed and the resultant expression is:
\begin{eqnarray}
p_{++;++}^{(0)}(t)&=&1-N(gt)^2+\sum_{n,n'=1}^\infty (-1)^{n+n'} \frac{(gt)^{2n+2n'}}{(2n)!(2n')!}\nn \\
&&\ \ \ \ \times N (N-1)...(N-n-n'+1),
\label{eq_numberstate1}
\end{eqnarray}
which is already different from the classical Rabi solution. The coherent photon dressing described in Section~\ref{sec_diagrammatic solution} no longer applies and one has to perform an infinite sum as in Eq.~(\ref{eq_numberstate1}).

\begin{figure}[htb]
\centering
\includegraphics[width=1.0\columnwidth]{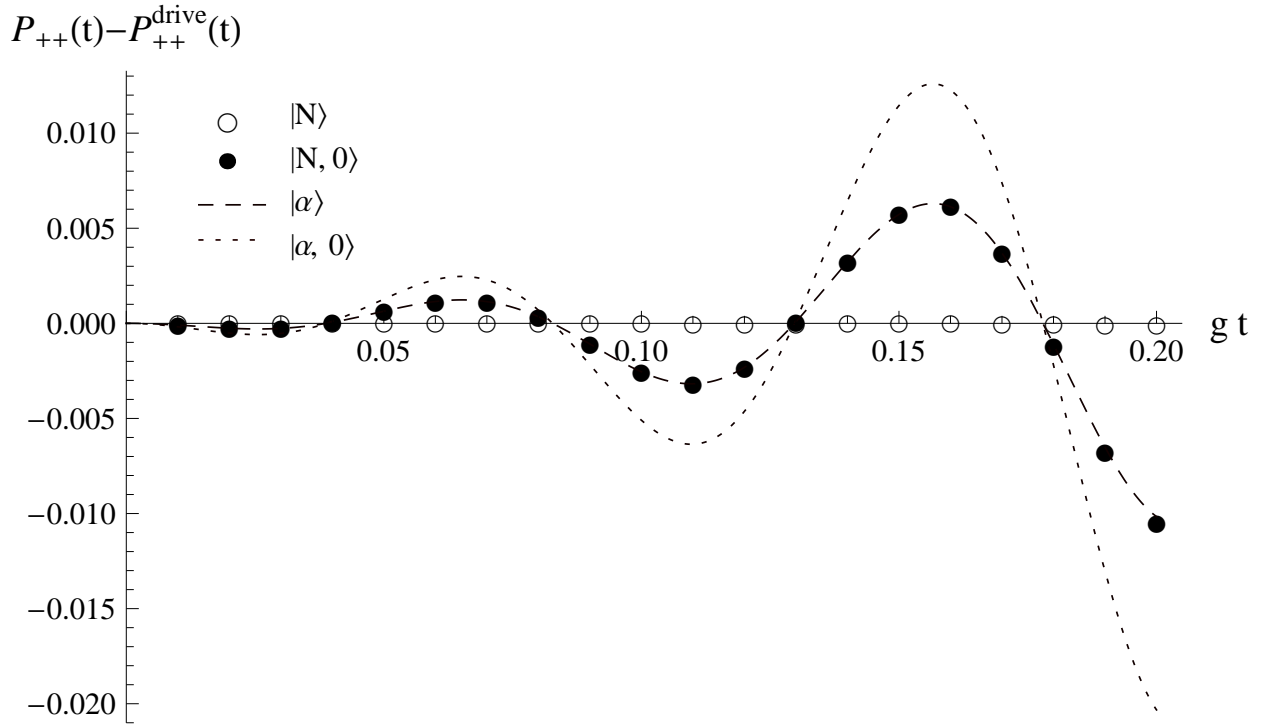}
\caption{Relaxation of an initially excited TLS under different initial photon states. $N=\left| \alpha \right|^2 =10^3$. The case with a non-dissipative single mode number state ($\left|N \right\rangle$) gains decoherence when a degenerate vacuum mode is added ($\left|N,0 \right\rangle$). In the small time region, the amount of dissipation is the same as that of a single mode coherent state ($\left|\alpha \right\rangle$).
 numberstate.eps}
\label{numberstate}
\end{figure}

The contraction line remains as a good small parameter. The first contraction terms can be calculated as in the way to obtain Eq.~(\ref{eq_numberstate1}). Fig.~\ref{numberstate} shows the relaxation, defined by $P_{++}(t)-P_{++}^{\text{drive}(t)}$, of a TLS driven by different initial photon states. For simplicity, we consider the vacuum mode to be degenerate with the $N$-mode, so that the contraction function is time independent. We define the driving term $P_{++}^{\text{drive}}(t)=\cos^2 (g\sqrt{N+1}\ t)$ and $\cos^2(g|\alpha| t)$ for the number and coherent states, respectively. In the case of a single mode number state $\left|N \right \rangle$, no relaxation is seen as expected. However, when we consider a number state with one degenerate vacuum mode $\left|N,0 \right \rangle$, dissipation is observed due to the interference between the control photon and the vacuum. The amount of decoherence increases linearly with the number of degenerate vacuum modes present. It is also interesting to observe that the decoherence of $\left|N,0 \right \rangle$ is equal to that of a single mode coherent state $\left|\alpha \right \rangle$ in the small time regime.

The diagrammatic approach serves as a first step towards a precise treatment of this multimode number state problem. The generalization to deal with states having a general photon number distribution remains as a challenge. The ME approximation schemes are even worse at accounting for the decoherence effect of this problem compared to their efficiencies in coherent state case. Following the discussion in Section~\ref{sec_comparison}, one can show that for an initial photon state $\left|N,0 \right \rangle$, the NZ ME leads to a non-dissipative oscillations of the TLS, where $P_{++}(t)$ can be negative even in the small time regime $gt \ll 1$. The TCL ME results in a Gaussian decay ($P_{++}^{TCL}(t) \approx [1+\exp\left(-2Ng^2 t^2 \right) ]/2$, for $N\gg 1$), where the Rabi oscillations are lost. Their failures again originate from the inability to capture the coupled dynamics between the control and the environment.

\section{Conclusion}\label{sec_conclusion}

The diagrammatic formalism provides a new platform to understand and compute the interaction dynamics among the quantum system, the control photon and the environment. Its high accuracy is attributed by the full account of the unitary evolution of the wavefunction of the whole system without any stochastic assumption. The revelation of underlying physical processes by the diagrammatic structures enables a perturbative computation scheme for systems under fast photon control versus slow relaxation, with well-defined error bounds. It also serves as an extension of the standard field theoretic techniques in quantum electrodynamics and many-body problems to systems far from equilibrium.

The ME approach has been a good phenomenology to provide qualitative understandings of open quantum systems. Given the recent advances in high precision quantum technologies, there is a need to reexamine and improve the accuracies of the existing theories. The diagrammatic solution indicates the absence of quantum interference between the control and the environment in current ME approximations. This general effect is not restricted to a TLS interacting with a single mode photon state. We believe that the diagrammatic results can motivate improvements of conventional ME studies and stimulate the developments of other novel quantum theories. There is hope that more quantum mechanical effects of open quantum systems will be uncovered in the future.

\section*{Acknowledgments}
This research was supported by the U.S. Army Research Office MURI award W911NF0910406.

\clearpage

\clearpage

\clearpage

\clearpage

\clearpage

\clearpage

\clearpage

\end{document}